\documentclass{svmult}
\usepackage{amssymb}

\usepackage{makeidx}     
\usepackage{graphicx}    
\usepackage{multicol}    

\makeindex             

\newcommand{\be}{\begin{equation}}
\newcommand{\ee}{\end{equation}}
\newcommand{\bea}{\begin{eqnarray}}
\newcommand{\eea}{\end{eqnarray}}
\newcommand{\beas}{\begin{eqnarray*}}
\newcommand{\eeas}{\end{eqnarray*}}

\newcommand{\treea}[1]{
\begin{picture}(10,4)(-5,-3)
\put(-2.5,-2){$\bullet #1$}
\end{picture}
}

\newcommand{\treeb}[2]{
\begin{picture}(10,10)(-5,-10)
\put(-2.5,-5){$\bullet #1$} \put(0,-4){\line(0,-1){6}}
\put(-2.5,-12){$\bullet #2$}
\end{picture}
}

\newcommand{\treeca}[3]{
\begin{picture}(10,10)(-5,-10)
\put(-2.5,-5){$\bullet #1$} \put(0,-4){\line(0,-1){6}}
\put(-2.5,-12){$\bullet #2$}
\put(0,-12){\line(0,-1){6}}\put(-2.5,-19){$\bullet #3$}
\end{picture}
}

\newcommand{\treecb}[3]{
\begin{picture}(20,10)(-10,-10)
\put(-2.5,-5){$\bullet #1$}\put(0,-3){\line(1,-2){5}}
\put(2,-14){$\bullet #2$}
\put(0,-3){\line(-1,-2){5}}\put(-12,-14){$#3 \bullet$}
\end{picture}
}

\newcommand{\treeda}[4]{
\begin{picture}(10,10)(-5,-10)
\put(-2.5,-5){$\bullet #1$} \put(0,-4){\line(0,-1){6}}
\put(-2.5,-12){$\bullet #2$}
\put(0,-12){\line(0,-1){6}}\put(-2.5,-19){$\bullet #3$}
\put(0,-18){\line(0,-1){6}}\put(-2.5,-25){$\bullet #4$}
\end{picture}
}

\newcommand{\treedb}[4]{
\begin{picture}(10,10)(-5,-10)
\put(-2.5,-5){$\bullet #1$} \put(0,-3){\line(1,-2){5}}
\put(0,-3){\line(-1,-2){5}}\put(-12,-15){$#2 \bullet$}
\put(2,-15){$\bullet #3$}
\put(5,-12){\line(0,-1){6}}\put(2,-21){$\bullet #4$}
\end{picture}
}

\newcommand{\treedc}[4]{
\begin{picture}(20,10)(-10,-10)
\put(-2.5,-5){$\bullet #1$}\put(0,-3){\line(0,-1){5}}
\put(-2.5,-11){$\bullet #2$}\put(0,-9){\line(1,-2){5}}
\put(2,-20){$\bullet #3$}
\put(0,-9){\line(-1,-2){5}}\put(-12,-20){$#4 \bullet$}
\end{picture}
}

\def\qedv{\;\raisebox{-1mm}{\includegraphics[height=4mm]{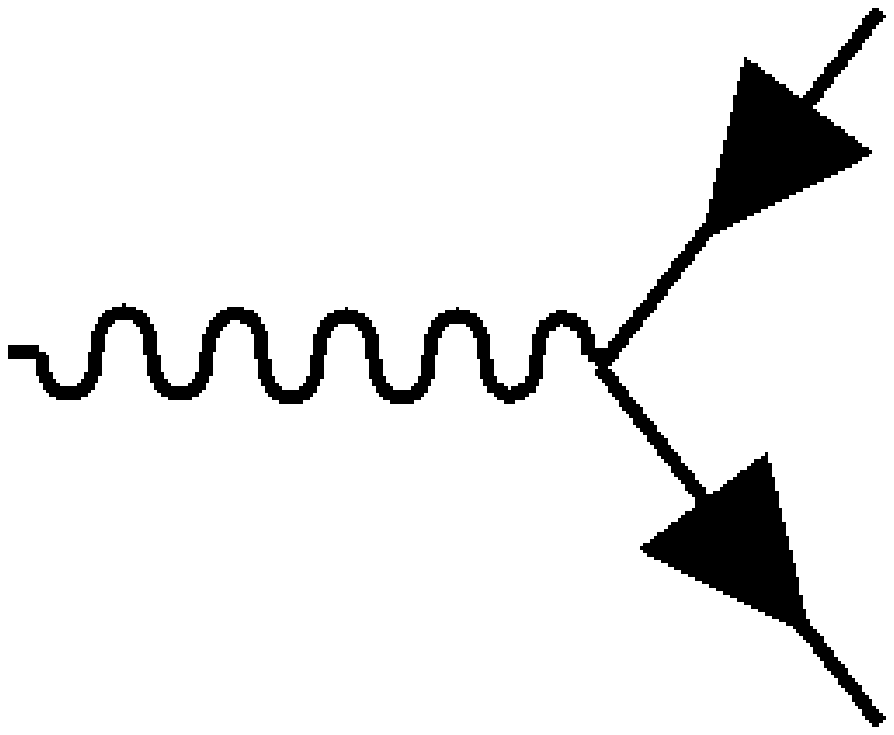}}\;}
\def\qedf{\;\raisebox{-0.5mm}{\includegraphics[height=2mm]{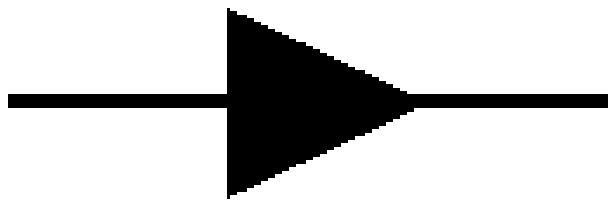}}\;}
\def\qedph{\;\raisebox{-0.5mm}{\includegraphics[height=2mm]{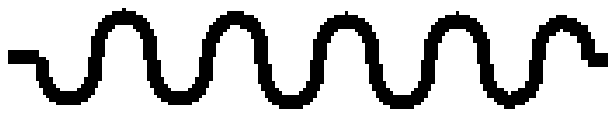}}\;}

\def\qcdfv{\;\raisebox{-1mm}{\includegraphics[height=4mm]{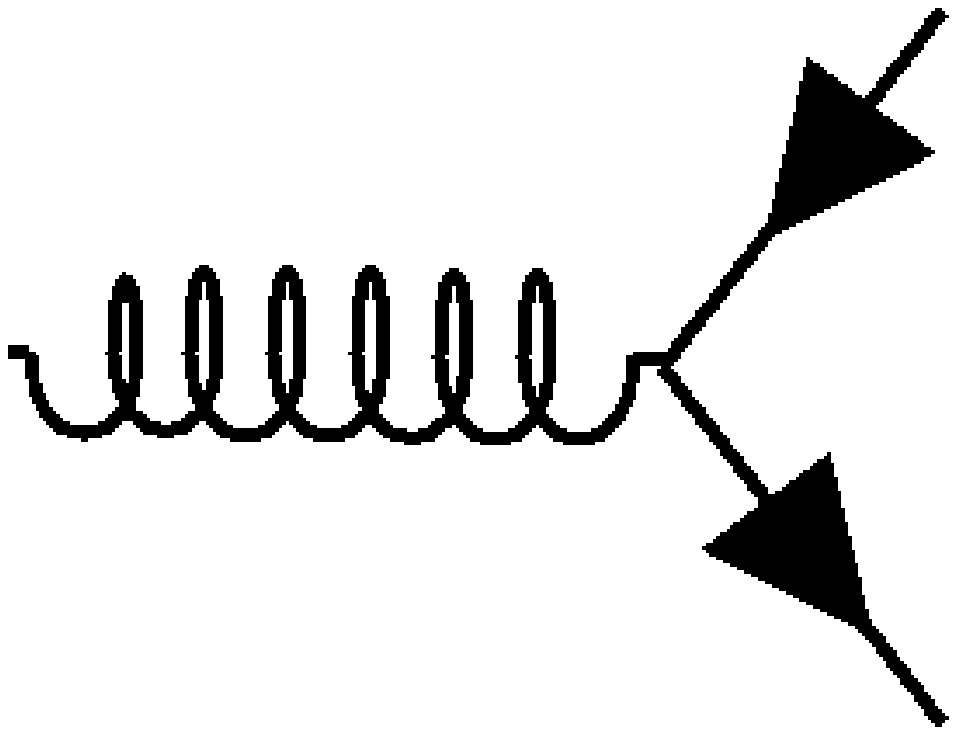}}\;}
\def\qcdgb{\;\raisebox{-0.5mm}{\includegraphics[height=2mm]{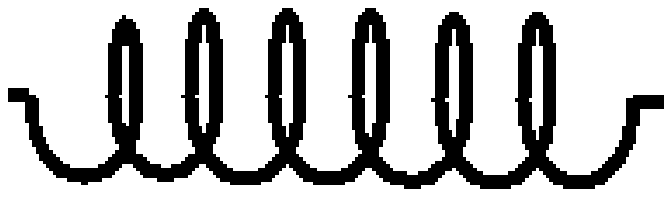}}\;}
\def\qcdgc{\;\raisebox{-1mm}{\includegraphics[height=4mm]{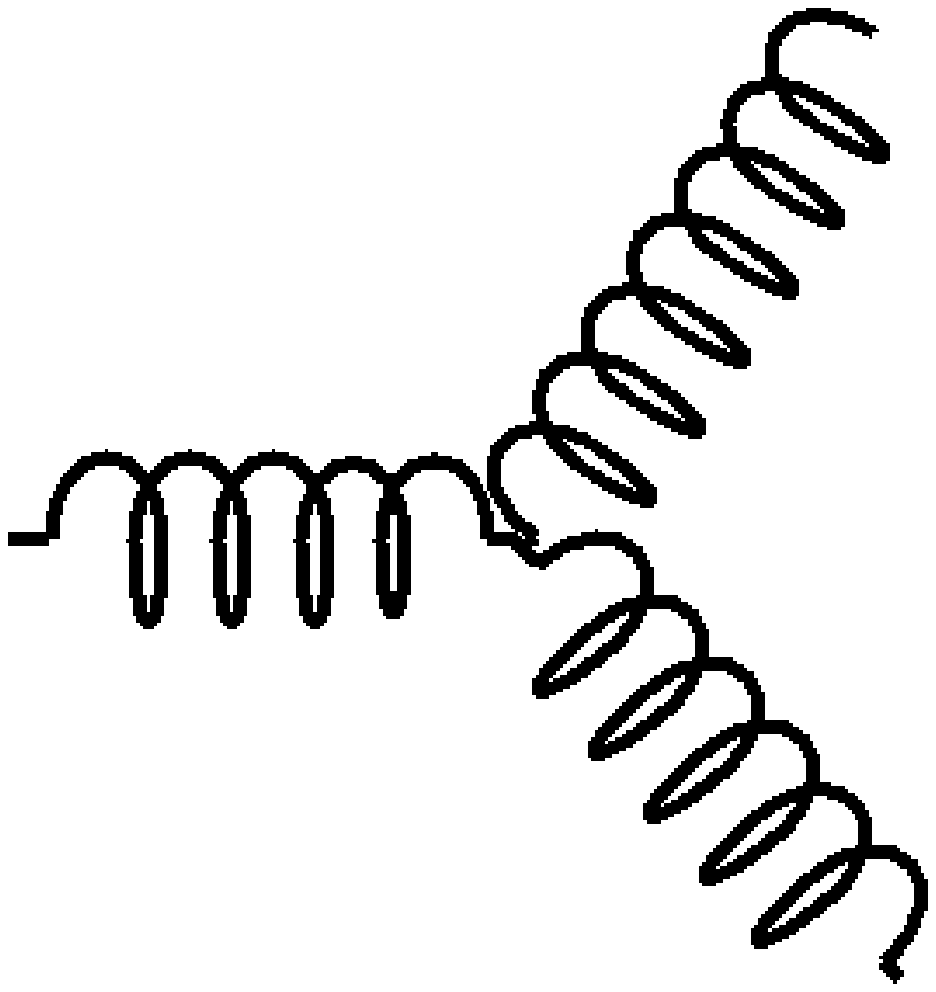}}\;}
\def\qcdgd{\;\raisebox{-1mm}{\includegraphics[height=4mm]{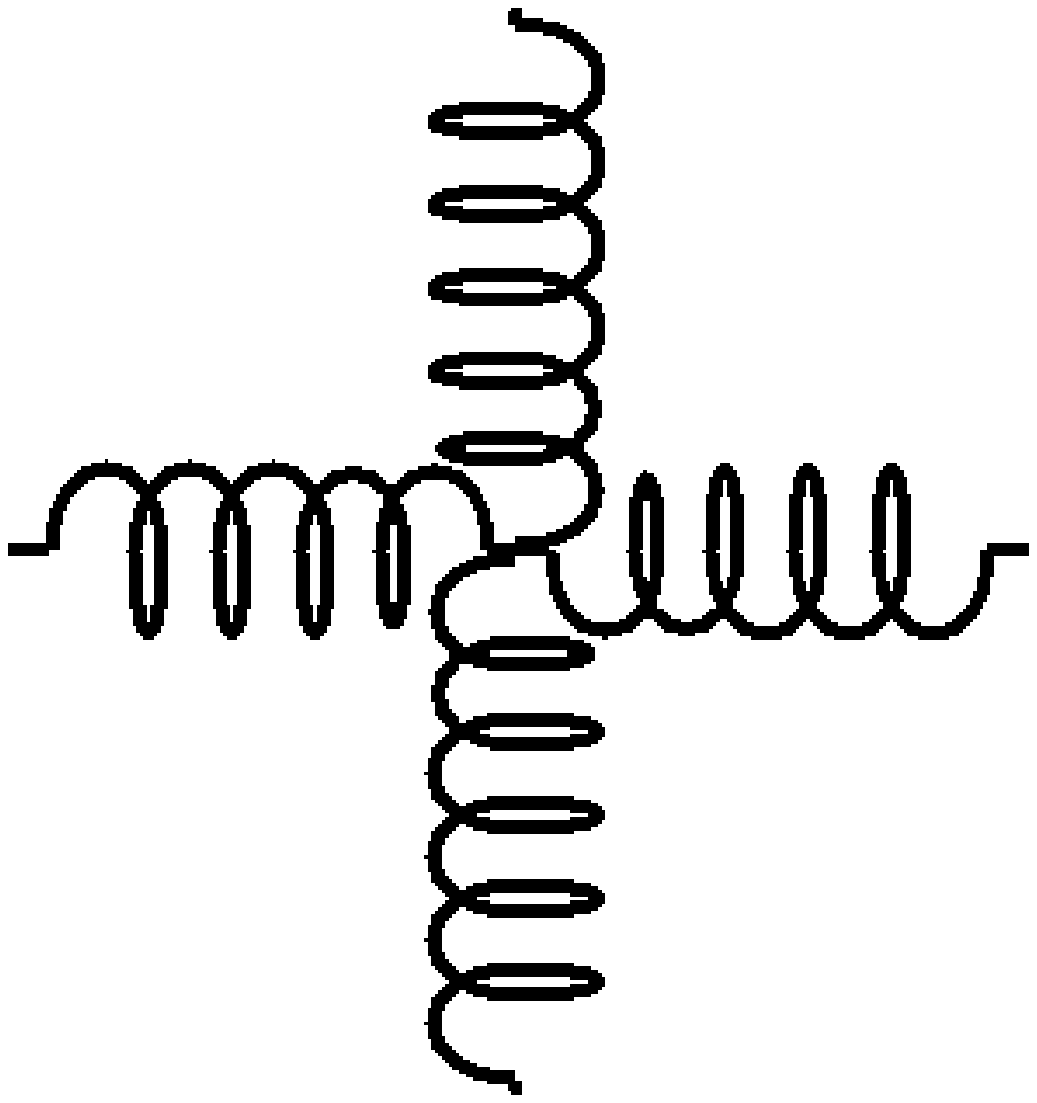}}\;}
\def\qcdghv{\;\raisebox{-1mm}{\includegraphics[height=5mm]{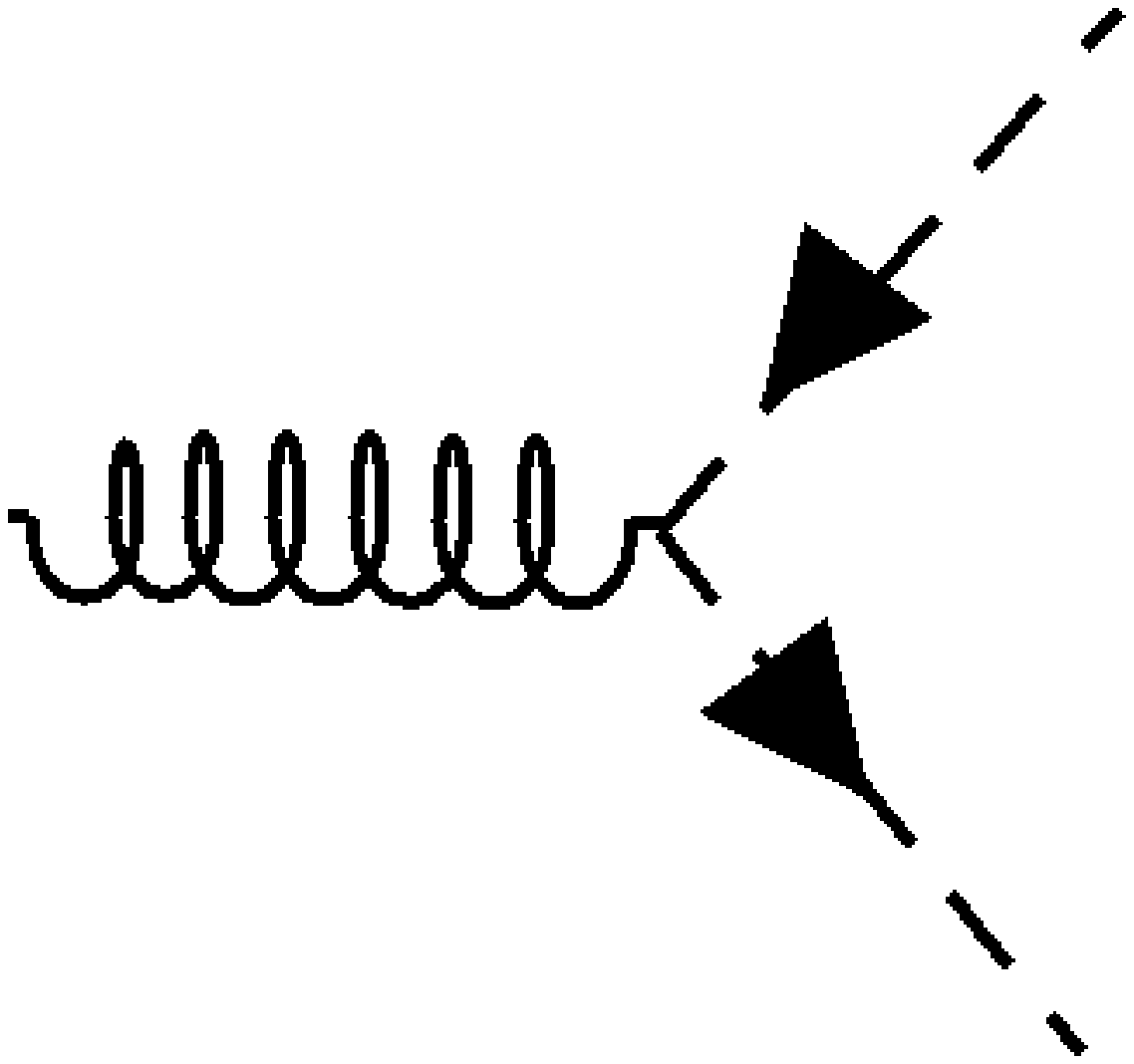}}\;}
\def\qcdgh{\;\raisebox{-0.5mm}{\includegraphics[height=2mm]{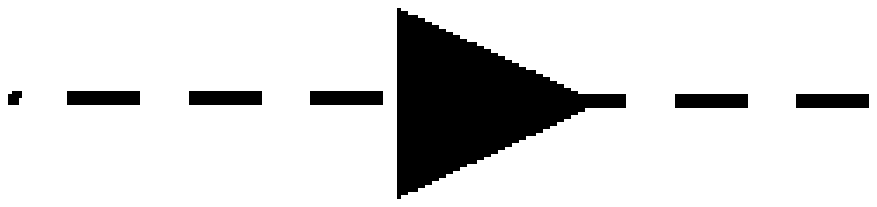}}\;}

\def\build#1_#2^#3{\mathrel{
\mathop{\kern 0pt#1}\limits_{#2}^{#3}}}

\begin{document}

\title*{Factorization in quantum field theory:  an exercise
in Hopf algebras and local singularities}
\titlerunning{Factorization in QFT}
\author{Dirk Kreimer\inst{}}
\institute{C.N.R.S.~at Institut des Hautes \'Etudes Scientifiques\\ 35
rte.~de Chartres, F91440 Bures-sur-Yvette\\
\texttt{kreimer@ihes.fr} }
%
%
\maketitle

{\small I discuss the role of Hochschild cohomology in Quantum
Field Theory with particular emphasis on Dyson--Schwinger
equations. Talk given at {\em Frontiers in Number Theory, Physics
and Geometry}; Les Houches, March 2003 [arXiv:hep-th/0306020].}

\section{Introduction}
This paper provides a designated introduction to the Hopf algebra
approach to renormalization having a specific goal in mind:  to
connect this approach in perturbative quantum field theory with
non-perturbative aspects, in particular with Dyson--Schwinger
equations (DSEs) and with the renormalization group (RG), with
particular emphasis given to  a proof of renormalizability based
on the Hochschild cohomology of the  Hopf algebra behind a
perturbative expansion.

To achieve this goal we will consider a Hopf algebra of decorated
rooted trees. In parallel work, we will treat the Feynman graph
algebras of quantum electrodynamics, non-abelian gauge theories
and the full Standard Model along similar lines \cite{DKnew}.

 There are various
reasons for starting with decorated rooted trees. One is that Hopf
algebra structures of such rooted trees play a prominent role also
in the study of polylogarithms
\cite{Gangl,Cartier,Goncharov,Zagier} and quite generally in the
analytic study of functions which appear in high-energy physics
\cite{Bern,Weinzierl}. Furthermore, the Hopf algebras of graphs
and decorated rooted trees are intimately related. Indeed,
resolving overlapping divergences into non-overlapping sectors
furnishes a homomorphism from the Feynman graph Hopf algebras to
Hopf algebras of decorated rooted trees
\cite{DK1,overl,RH1,CKnew}.  Thus the study of decorated rooted
trees is by no means a severe restriction of the problem, but
allows for the introduction of simplified models which still
capture the crucial features of the renormalization problem in a
pedagogical manner.\footnote{Furthermore, the structure of the
Dyson--Schwinger equations in gauge theories eliminates
overlapping divergences altogether upon use of gauge invariance
\cite{BJW,BDK}.}

In particular we are interested to understand how the structure
maps of a Hopf algebra allow to illuminate the structure of
quantum field theory. We will first review the transition from
unrenormalized to renormalized amplitudes
\cite{DK1,chen,BD1,CK1,RH1,BK} and investigate how the Hochschild
cohomology of the Hopf algebra of a perturbative expansion
directly leads to a renormalization proof.

We then study Dyson--Schwinger equations  for rooted trees and
show again how the Hochschild cohomology explains the form
invariance of these equations under the transition from the
unrenormalized to the renormalized equations. For the Hopf
algebras apparent in a perturbative expansion, this transition is
equivalent to the transition from the action to the bare action,
as the study of Dyson--Schwinger equations is equivalent to the
study of the corresponding generating functionals
\cite{Rivers,Predrag}.

We then show how the structure of these equations leads to a
combinatorial factorization into primitives of the Hopf algebra.
While this is easy to achieve for the examples studied here, it is
subtly related to the Ward--Takahashi and Slavnov--Taylor
identities in the case of abelian and non-abelian quantum gauge
field theories. Here is not the space to provide a detailed
discussion of factorization in these theories, but at the end of
the paper we comment on recent results of \cite{DKnew} concerning
the relation between factorization and gauge symmetry. Indeed, the
combinatorial factorization establishes a commutative associative
product $\vee$ from one-particle irreducible (1PI) graphs to 1PI
graphs in the Hopf algebra of 1PI graphs. In general, this product
is non-integral \cite{banz}: \be \Gamma_1\vee\Gamma_2=0
\not\Rightarrow \Gamma_1=0\; \mbox{or}\; \Gamma_2=0,\ee but the
failure  can be attributed to the one-loop graphs generated from a
single closed fermion loop with a suitable number of external
background gauge fields coupled. This fermion determinant is the
starting point in \cite{DKnew} for an understanding of gauge
symmetries based on an investigation of the structure of the ring
of graph insertions.

Having a commutative ring  at hand of 1PI graphs, or, here, of
decorated rooted trees, we can ask how the evaluation of a product
of 1PI graphs, or trees,  compares with the product of the
evaluations. In answering this question, it seems to me, serious
progress can be made in our understanding of field theory. Indeed,
the integrals which appear in Dyson--Schwinger equations or in the
perturbative expansion of field theory are of a distinguished
kind: they provide a class of functions which is self-similar
under the required integrations. The asymptotics of the integral
can be predicted from the asymptotics of the integrand, as already
stressed by previous authors \cite{W}. It is this self-similarity
which makes the Dyson--Schwinger equations consistent with the
renormalization group. Again, a detailed study has to be given
elsewhere but a few comments are scattered in the present paper.

   We will now outline this program in some detail, and
then first turn to a rich class of toy models to exhibit many of
the involved concepts. This serves as a training ground for our
ideas.  As announced, these toy models are based on a Hopf algebra
of decorated rooted trees, with only symbolically specified
decorations. We provide toy Feynman rules which suffer from short
distance singularities. Each genuine quantum field theory is
distinguished from this toy case by the mere fact that the
calculation of the decorations is analytically harder than what
confronts the reader later on. Any perturbative quantum field
theory (pQFT)   provides a Hopf algebra structure isomorphic to
the models below, for a suitably defined set of decorations,
through its skeleton graphs.

Unfortunately, the calculation of higher loop order skeletons is
beyond the present analytical skill. Most fascinatingly though, up
to six loops, they provide multiple zeta values galore
\cite{DKbook}, a main subject of our school
\cite{Bern,Weinzierl,Zagier,Cartier}. At higher loops, they might
even provide periods outside this class, an open research question
in its own right \cite{Bel}.

 Nevertheless, there is still much to be learned
about how the underlying skeleton diagrams combine in quantum
field theory. Ultimately, we claim that the Hopf- and Lie algebra
structures of 1PI graphs are sufficiently strong to reduce quantum
field theory to a purely analytical challenge: the explanation of
relations between two-particle irreducible (2PI) graphs which will
necessitate the considerations of higher Legendre transforms. This
is not the purpose of the present paper, but a clear task for the
future:  while the renormalization problem of 1PI graphs is
captured by the algebraic structures of 1PI graphs, the analytic
challenge is not: Rosner's cancellation of transcendentals in the
$\beta$ function of quenched QED \cite{Rosner,BDK}, Cvitanovic's
observation of hints towards non-combinatorial growth of
perturbative QED \cite{Cvi} and the observation of (modified) four
term relations between graphs \cite{4term} all establish relations
between 2PI skeleton graphs which are primitives in the Hopf
algebra of 1PI graphs. In this sense, the considerations started
in this paper aim to emphasize where the true problem of QFT lies:
in the understanding of the analytic relations between
renormalization primitive graphs. The factorizations into Hopf
algebra primitives of the perturbation expansion studied here
generalizes the shuffle identity on generalized polylogarithms,
which comes, for the latter, from studying the very simple
integral representations as iterated integrals. A second source of
relations comes from studying the sum representations. The
corresponding relations between Feynman diagrams have not yet been
found, but the above quoted results are, to my mind, a strong hint
towards their existence. Alas, the lack of understanding of these
relations is the major conceptual challenge which stops us from
understanding QFT in four dimensions. All else is taken care of by
the algebraic structures of 1PI graphs.

The Hopf algebra of decorated rooted trees is an adequate training
ground for QFT, where the focus is on the understanding of the
renormalization problem and the factorization of 1PI graphs into
graphs which are primitive with respect to the Hopf algebra
coproduct.

Hence the  program which we want to carry out in the following
consists of a series of steps which can be set up in any QFT,
while in this paper we will utilize the fact that they can be set
up in a much wider context. When one considers DSE, one usually
obtains them as the quantum equations of motion of some Lagrangian
field theory using some generating functional technology in the
path integral. Now the DSEs for 1PI Green functions can all be
written in the form \be \Gamma^{\underline{n}}= 1 +
\sum_{\gamma\in H_L^{[1]} \atop {\rm res}(\gamma)=\underline{n}}
\frac{\alpha^{\vert\gamma\vert}}{{\rm Sym}(\gamma)}
B_+^\gamma(X_{\cal R}^\gamma), \ee  where the $B_+^\gamma$ are
Hochschild closed one-cocycles of the Hopf algebra of Feynman
graphs indexed by Hopf algebra primitives $\gamma$ with external
legs $\underline{n}$, and $X_{\cal R}^\gamma$ is a monomial in
superficially divergent Green functions which dress the internal
vertices and edges of $\gamma$. We quote this result from
\cite{DKnew} to which we refer the reader for details. It allows
to obtain the quantum equations of motion, the DSEs for 1PI Green
functions, without any reference to actions, Lagrangians or path
integrals, but merely from the representation theory of the
Poincar\'e group for free fields.

Motivated by this fact we will from now on call any equation of
the form \be X=1+\alpha B_+(X^k),\ee with $B_+$ a closed
Hochschild one-cocycle, a combinatorial Dyson--Schwinger equation.

Thus in this paper  we choose as a first Hopf algebra to study the
one of  decorated rooted trees, without specifying a particular
QFT. The decorations play the role of the skeleton diagrams
$\gamma$ above, indexing the set of closed Hochschild one-cocycles
and the primitives of the Hopf algebra.

In general, this motivates an approach to quantum field theory
which is utterly based on the Hopf and Lie algebra structures of
graphs. Let us discuss the  steps which we would have to follow in
such an approach.
\subsection{Determination of $H$}
The first step aims at finding the Hopf algebra suitable for the
description of a  chosen QFT. For such a QFT consider the set of
Feynman graphs corresponding to its perturbative expansion close
to its free Gaussian functional integral. Identify the
one-particle irreducible (1PI) diagrams. Identify all vertices and
propagators in them and define a pre-Lie product on 1PI graphs by
using the possibility to replace a local vertex by a vertex
correction graph, or, for internal edges, by replacing a free
propagator by a self-energy. For any local QFT this defines a
pre-Lie algebra of graph insertions \cite{CKnew}. For a
renormalizable theory, the corresponding Lie algebra will be
non-trivial for only a finite number of types of 1PI graphs
(self-energies, vertex-corrections) corresponding to the
superficially divergent graphs, while the superficially convergent
ones provide a semi-direct product with a trivial abelian factor
\cite{CK1}.

The combinatorial graded pre-Lie algebra so obtained \cite{CKnew}
provides not only a Lie-algebra ${\cal L}$, but a commutative
graded Hopf algebra $H$ as the dual of its universal enveloping
algebra ${\cal U(L)}$, which is not cocommutative if ${\cal L}$
was non-abelian. Dually one hence obtains a commutative but
non-cocommutative Hopf algebra $H$ which underlies the forest
formula of renormalization \cite{DK1,overl,chen,CK1}.
\subsection{Character of $H$} For a so-determined Hopf algebra $H
= H (m,E,\bar e,\Delta ,S)$, a Hopf algebra with multiplication
$m$, unit $e$ with unit map $E:{\mathbb Q}\to H$, $q\to qe$, with
counit $\bar e$, coproduct $\Delta$ and antipode $S$, $S^2 = e$,
we immediately have at our disposal the group of characters $G=
G(H)$ which are multiplicative maps from $H$ to some target ring
$V$. This group contains a distinguished element:
 the Feynman
rules $\varphi$ are indeed a very special  character in $G$. They
will typically suffer from short-distance singularities, and the
character $\varphi$ will correspondingly reflect these
singularities. This can happen in various ways depending on the
chosen target space $V$. We will here typically take $V$ to be the
ring of Laurent polynomials in some indeterminate $z$ with poles
of finite orders for each finite Hopf algebra element, and design
Feynman rules so as to reproduce all salient features of QFT.

As $\varphi : H \rightarrow V$, with $V$ a ring, with
multiplication $m_V$, we can introduce the group law \be \varphi *
\psi =  m_V \circ (\varphi \otimes \psi) \circ \Delta \, , \ee and
use it to define a new character \be S_R^{\phi} * \phi \in G \, ,
\ee where $S_R^{\phi} \in G$ twists $\phi \circ S$ and furnishes
the counterterm of $\phi (\Gamma)$, $\forall \, \Gamma \in H$,
while $ S_R^{\phi}
* \phi (\Gamma)$ corresponds to the renormalized contribution of $\Gamma$
\cite{DK1,chen,overl,CK1}. $S_R^\phi$ depends on the Feynman rules
$\phi:\;H\to V$ and the chosen re\-nor\-ma\-li\-za\-tion scheme
$R:\; V\to V$. It is given by \be  S_R^\phi=-R\left[m_V\circ
(S_R^\phi \otimes \phi)\circ ({\rm id}_H\otimes
P)\circ\Delta\right]\; ,\ee where $R$ is supposed to be a
Rota-Baxter operator in $V$, and the projector into the
augmentation ideal  $P:H\to H$ is given by $P={\rm id}-E\circ
\bar{e}$.

The $\bar R$ operation of Bogoliubov is then given by \be  \bar
\phi:= \left[m_V\circ (S_R^\phi \otimes \phi)\circ ({\rm
id}_H\otimes P)\circ\Delta\right]\; ,\ee  and \be
S_R^\phi\star\phi\equiv m_V\circ (S_R^\phi\otimes
\phi)\circ\Delta=\bar\phi+S_R^\phi=({\rm id}_H-R)(\bar\phi )\ee is
the renormalized contribution. Note that this second step  has
been established for all perturbative quantum field theories
combining the results of \cite{DK1,overl,RH1,CKnew,CK1}. These
papers are rather abstract and will be complemented by explicit
formulas for the practitioner of gauge theories in forthcoming
work.

\subsection{Locality from $H$} The third step aims to show that locality of
counterterms is utterly determined by the Hochschild cohomology of
Hopf algebras. Again, we can dispense of the existence of an
underlying Lagrangian and derive this crucial feature from the
Hochschild cohomology of $H$. This cohomology is universally
described in \cite{CK1}, see also \cite{Foissy}.  What we are
considering are spaces ${\cal H}^{(n)}$ of maps from the Hopf
algebra into its own $n$-fold tensor product, \be {\cal
H}^{(n)}\ni \psi\Leftrightarrow \psi: H\to H^{\otimes n} \ee  and
an operator \be b:\; {\cal H}^{(n)}\to {\cal H}^{(n+1)}\ee  which
squares to zero: $b^2=0$. We have for $\psi\in {\cal H}^{(1)}$ \be
(b\psi)(a)=\psi(a)\otimes e-\Delta(\psi(a))+({\rm id}_H\otimes
\psi)\Delta(a)\ee  and in general \be
(b\psi)(a)=(-1)^{n+1}\psi(a)\otimes e +\sum_{j=1}^n (-1)^j
\Delta_{(j)}\left( \psi(a)\right)+({\rm
id}\otimes\psi)\Delta(a),\ee where $\Delta_{(j)}:H^{\otimes n}\to
H^{\otimes n+1} $ applies the coproduct in the $j$-th slot of
$\psi(a)\in H^{\otimes n}$.

For all the Hopf algebras considered here and in future work on
QFT, the Hochschild cohomology is rather simple: it is trivial in
degree $n>1$, so that the only non-trivial elements in the
cohomology are the maps from $H\to H$ which fulfil the above
equation and are non-exact. In QFT these maps are given by maps
$B_+^\gamma$, indexed by primitive graphs $\gamma$, an easy
consequence of \cite{CK1,Foissy} extensively used in
\cite{DKnew}.

Locality of counterterms and finiteness of renormalized quantities
follow from the Hochschild properties of $H$: the Feynman graph is
in the image of a closed Hochschild one cocycle $B_+^{\gamma}$, $b
\, B_+^{\gamma} =  0$, i.e. \be \Delta \circ B_+^{\gamma} (X) =
B_+^{\gamma} (X) \otimes e + ({\rm id} \otimes B_+^{\gamma}) \circ
\Delta (X) \, , \ee and this equation suffices to prove the above
properties by a recursion over the augmentation degree of $H$.
This is a new result: it is the underlying Hochschild cohomology
of the Hopf algebra $H$ of the perturbative expansion which allows
to provide renormalization by local counterterms. The general case
is studied in \cite{DKnew}, but we will study this result in
detail for rooted tree algebras below. This result is again valid
due to the benign properties of Feynman integrals: we urgently
need Weinberg's asymptotic theorem which ensures that an
integrand, overall convergent by powercounting and free of
subdivergences, can actually be integrated \cite{W}.

\subsection{Combinatorial DSEs from Hochschild cohomology}
Having understood the mechanism which achieves locality
step by step in the perturbative expansion, one can ask for more:
how does this mechanism fare in the quantum equations of motion?
So we next turn to the Dyson--Schwinger equations.

 As mentioned before, they typically are of the form
\be \Gamma^{\underline{n}}= 1 + \sum_{\gamma\in H_L^{[1]} \atop
{\rm res}(\gamma)=\underline{n}}
\frac{\alpha^{\vert\gamma\vert}}{{\rm Sym}(\gamma)}
B_+^\gamma(X_{\cal R}^\gamma)=1+\sum_{\Gamma\in H_L\atop {\rm
res}(\Gamma)=\underline{n}}\frac{\alpha^{\vert\Gamma\vert}\Gamma}{{\rm
Sym}(\Gamma)}\; , \ee
 where the first sum is over a finite (or
countable) set of Hopf algebra primitives $\gamma$, \be
\Delta(\gamma)=\gamma\otimes e + e \otimes \gamma,\ee  indexing
the closed Hochschild one-cocycles $B_+^{\gamma}$ above, while the
second sum is over all one-particle irreducible graphs
contributing to the desired Green function, all weighted by their
symmetry factors. The equality is non-trivial and needs proof
\cite{DKnew}.  Here, $\Gamma^{\underline{n}}$ is to be regarded as
 a formal series
\be  \Gamma^{\underline{n}}=1+\sum_{k\geq 1} c_k^{\underline{n}}
\alpha^k, \;c_k^{\underline{n}}\in H.\ee

\begin{figure}
\center
\includegraphics[height=2cm]{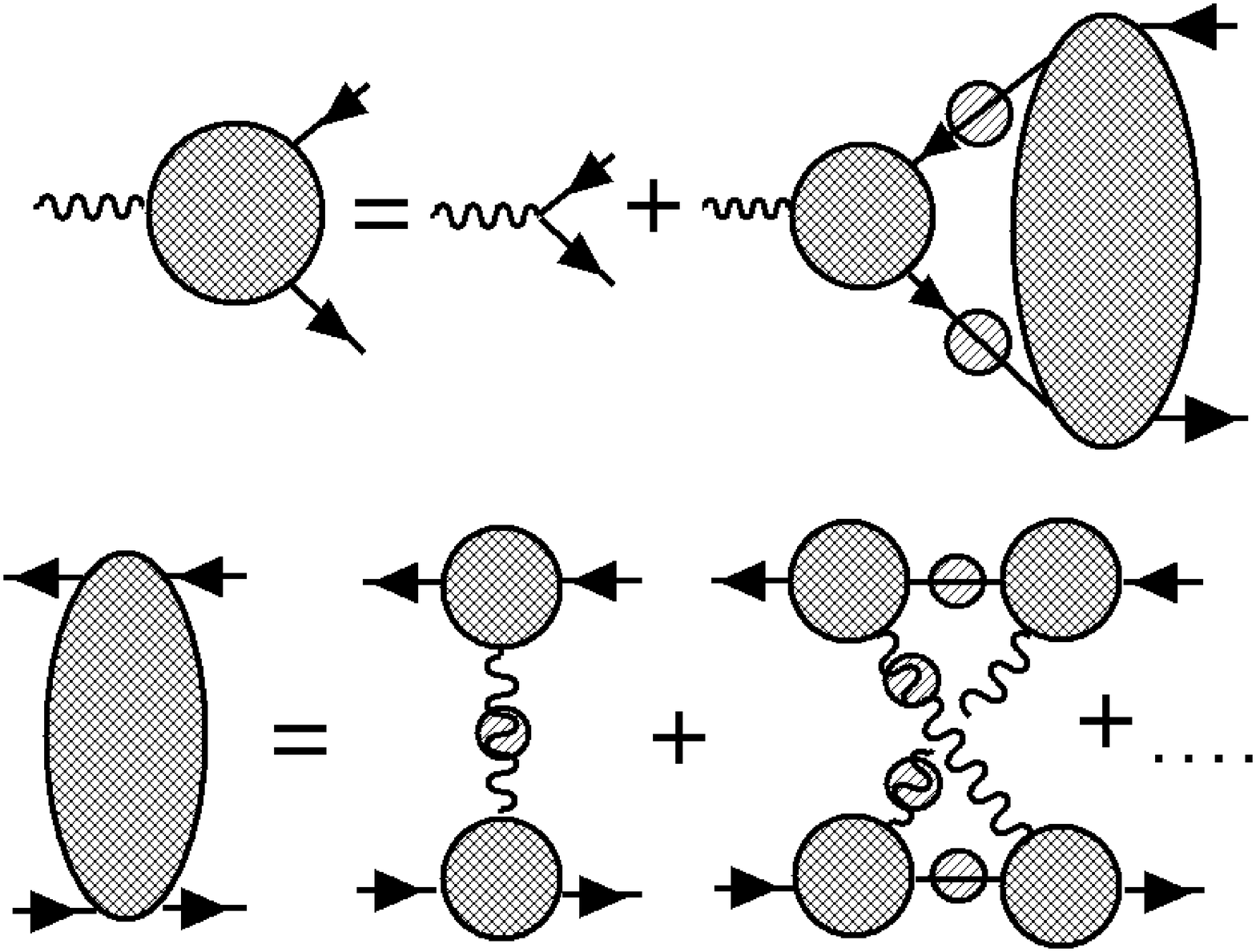}\includegraphics[height=1cm]{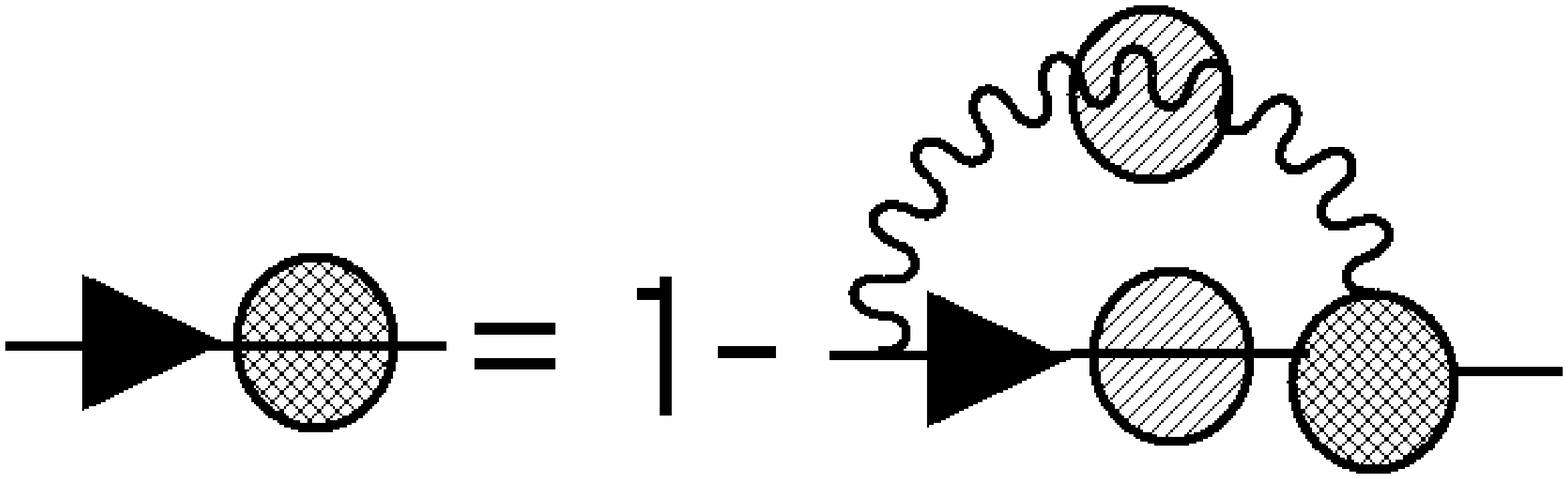}
\includegraphics[height=1cm]{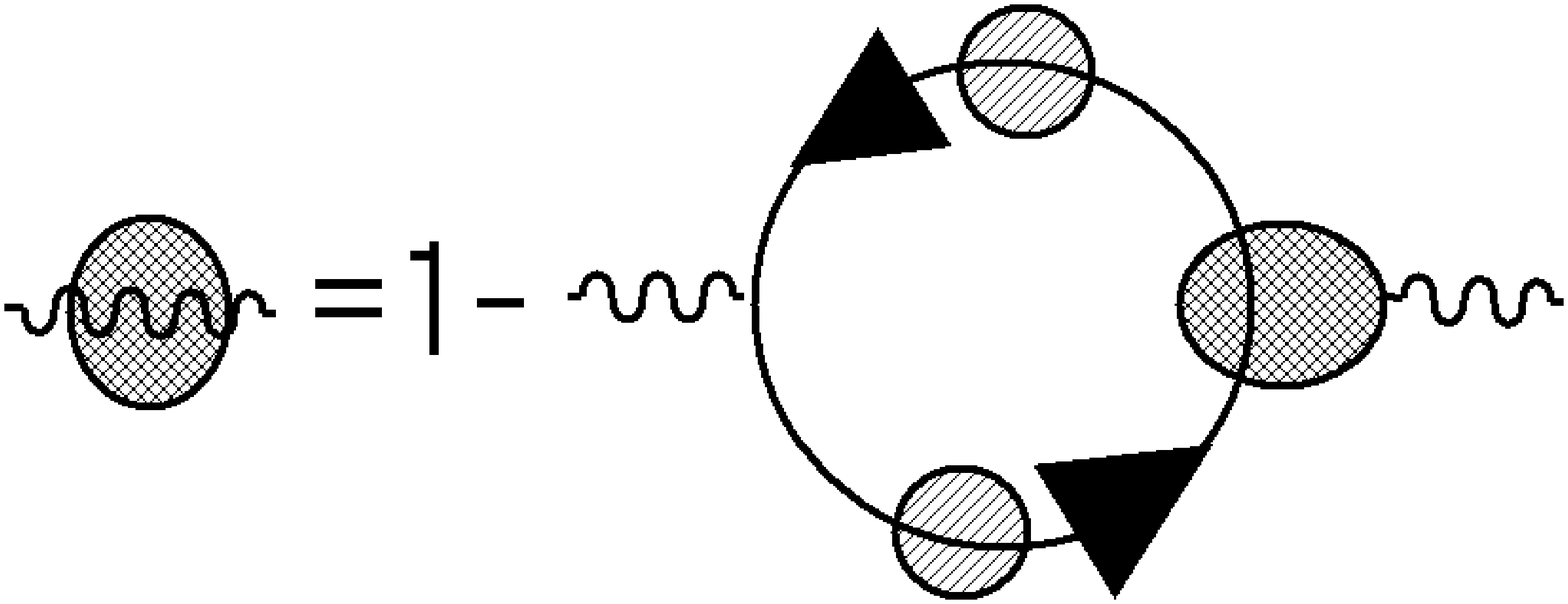}
\caption{The Dyson Schwinger equation for the QED vertex and
inverse propagator functions.
Expanding the $e^+ e^-$ scattering
kernel, and closing it by a 1PI vertex function, gives the terms
generated by the $B_+^\gamma$ operator: Replacing each blob by a
bare vertex or propagator gives the primitive graphs $\gamma$ for
the vertex function. Solving that DSE for the bare vertex and
inserting it into the equation for the inverse propagator function
gives their skeletons by standard methods \cite{BJW,BDK}.}
\label{fig:1}       
\end{figure}

 Typically, this is all summarized in graphical form as in
Fig.(\ref{fig:1}), which gives the DSE for the unrenormalized
Green  functions of massless QED as an example (restricting
ourselves to the set of superficially divergent Green functions,
ie.~$\underline{n}\in {\cal R}_{\rm QED}\equiv\{
\qedv,\qedf,\qedph \} $). In our terminology this QED system reads
for renormalized functions: {\small \beas \Gamma^{\qedv}_R & = &
Z^{\qedv}+\!\!\!\!\!\!\!\!\sum_{\gamma\in H_L^{[1]}\atop {\rm
res}(\gamma)=\qedv}\!\!\!\!\frac{\alpha^{\vert\gamma\vert}}{{\rm
Sym}(\gamma)}B_+^\gamma\left([\Gamma^{\qedv}_R]^{n^\gamma_{\qedv}}/[\Gamma^{\qedf}_R]^{n^\gamma_{\qedf}}/[\Gamma^{\qedph}_R]^{n^\gamma_{\qedph}}\!\!\right)\\
\Gamma^{\qedf}_R & = & Z^{\qedf}+\!\!\!\!\!\!\!\!\sum_{\gamma\in
H_L^{[1]}\atop {\rm
res}(\gamma)=\qedf}\!\!\!\!\frac{\alpha^{\vert\gamma\vert}}{{\rm
Sym}(\gamma)}B_+^\gamma\left([\Gamma^{\qedv}_R]^{n^\gamma_{\qedv}}/[\Gamma^{\qedf}_R]^{n^\gamma_{\qedf}}/[\Gamma^{\qedph}_R]^{n^\gamma_{\qedph}}\!\!\right)\\
\Gamma^{\qedph}_R & = &
Z^{\qedph}+\!\!\!\!\!\!\!\!\!\sum_{\gamma\in H_L^{[1]}\atop {\rm
res}(\gamma)=\qedph}\!\!\!\!\!\!\frac{\alpha^{\vert\gamma\vert}}{{\rm
Sym}(\gamma)}B_+^\gamma\left([\Gamma^{\qedv}_R]^{n^\gamma_{\qedv}}/[\Gamma^{\qedf}_R]^{n^\gamma_{\qedf}}/[\Gamma^{\qedph}_R]^{n^\gamma_{\qedph}}\!\!\!\right)\eeas}
where the integers {\small
$$n^\gamma_{\qedv},\;n^\gamma_{\qedf},\;n^\gamma_{\qedph}$$} count
the numbers of internal vertices, fermion lines and photon lines
in $\gamma$, and the $B_+^\gamma$ operator inserts the
corresponding Green functions into $\gamma$, corresponding to the
blobs in  figure (\ref{fig:1}). The unrenormalized equations are
obtaind by omitting the subscript $R$ at $\Gamma^{\ldots}_R$ and
setting $Z^{\ldots}$ to unity. The usual integral equations are
obtained by evaluation both sides of the system by the Feynman
rules.\footnote{The system is redundant, as we made no use of the
Ward identity. Also, the unrenormalized system is normalized so
that the rhs starts with unity, implying an expansion of inverse
propagators in the external momentum up to their superficial
degree of divergence. This creates their skeleton diagrams
\cite{BJW,BDK}.}
The form invariance in the transition from the unrenormalized to
the renormalized Green functions directly follows from the fact
that the equation for the series $\Gamma^{\underline{n}}$ is in
its non-trivial part in the image of closed Hochschild
one-cocycles $B_+^\gamma$. It is this fact which ensures that a
local Z-factor is sufficient to render the theory finite. The fact
that the rhs of a DSE is Hochschild closed  ensures the form
invariance of the quantum equations of motion in the transition
from the unrenormalized to the renormalized Green functions, as
indeed in the renormalized system the Hochschild closed
one-cocycle acts only on renormalized functions. We will exemplify
this for rooted trees below.

\subsection{Factorization} Such systems of DSEs can be factorized.
The factorization is based on a commutative associative product on
one-particle irreducible (1PI) graphs in the Hopf algebra, which
maps 1PI graphs to 1PI graphs. We will do this below with
considerable ease for the corresponding product on rooted trees.
For Feynman graphs, one confronts the problem that the product can
be non-integral \cite{banz}. A detailed discussion of the relation
of this failure to the requirements of identities between Green
functions in the case of gauge theories is given in \cite{DKnew},
a short discussion appended at the end of this paper.

\subsection{Analytic factorization and the RG}
In the final step, we pose the question: how relates
the evaluation of the product to a product of the evaluations?

That can be carried out in earnest only in the context of a true
QFT \cite{DKnew}  - there is no exact RG equation available for
our toy model of  decorated rooted trees. So we will not carry out
this step here, but only include  in the discussion  at the end of
the paper an argument why a RG equation is needed for this step.

\section{Locality and Hochschild cohomology}
The first result we want to exhibit in some detail is the close
connection between the Hochschild cohomology of a Hopf algebra and
the possibility to obtain local counter\-terms.

We will first study the familiar Hopf algebra of non-planar
decorated rooted trees. We will invent toy Feynman rules for it
such that we have a non-trivial renormalization problem. Then we
will show how the structure maps of the Hopf algebra precisely
allow to construct local counterterms and finite renormalized
amplitudes thanks to the fact that each non-trivial Hopf algebra
element is in the image of a closed Hochschild one-cocycle.

\subsection{The Hopf algebra of decorated rooted trees}
To study the connection between renormalization and Hochschild
cohomology in a most comprehensive manner we thus introduce the
Hopf algebra of decorated rooted trees. Let ${\rm Dec}$ be a
(countable) set of decorations, and $H=H({\rm Dec})$ be the Hopf
algebra of decorated rooted trees (non-planar). For any such tree
$T$ we let $T^{[0]}$ be the set of its vertices and $T^{[1]}$ be
the set of its edges. To each vertex $v\in T^{[0]}$ there is
assigned a decoration ${\rm dec}(v)\in {\rm Dec}$.

For $T_1 , T_2$ in $H$, we let their disjoint union be the
product, we write $e$ for the unit element in $H$, and define the
counit by \be \bar e (e) =  1 \, , \quad \bar e (X) =  0 \
\hbox{else} . \ee We write $P:H\to H, P= {\rm id}_H-E\circ\bar
e\;$ for the projection into the augmentation ideal of $H$.

The coproduct is given by \be \Delta [e] =  e \otimes e \, , \
\Delta [T_1 \ldots T_k] =  \Delta [T_1] \ldots \Delta [T_k] \, ,
\ee \be \Delta [T] =  T \otimes e + e \otimes T + \sum_{{{{\rm
adm} \atop {\rm cuts}} \atop C}} P^C (T) \otimes R^C (T) \, , \ee
as in \cite{CK1}. The introduction of decorations does not require
any changes, apart from the fact that the operators $B_+$ are now
indexed by the decorations. We have, as $T =  B_+^c (X)$, for some
$X\in H$ and $c\in{\rm Dec}$,
 \be
\Delta [T] =  T \otimes e + [{\rm id} \otimes B_+^c] \, \Delta [X]
\, . \ee The antipode is given by $S(e)=e$ and \be S [B_+^c (X)] =
- B_+^c (X) -m \circ [S\!\circ\!P\otimes B_+^c] \circ \Delta [X]
\, . \label{S}\ee

A distinguished role is played by the primitive elements
$\treea{c}\;$,  $\forall c\in {\rm Dec}$, with \be
\Delta(\treea{c}\;)=\treea{c}\;\otimes e + e \otimes \treea{c}\;
.\ee

Let now $G$ be the group of characters of $H$, $\varphi \in G
\Leftrightarrow \varphi : H \rightarrow V$, $\varphi (T_1 T_2) =
\varphi (T_1) \, \varphi (T_2)$, with $V$ a suitable ring. Feynman
rules provide such characters for the Hopf algebras of QFT, and we
will now provide a character for the Hopf algebra of rooted trees
which mimicks the renormalization problem faithfully.

\subsection{The toy Feynman rule}
We choose $V$ to be the ring of Laurent series with poles of
finite order.  To understand the mechanism of renormalization in
an analytically simple case we define toy Feynman rules using
dimensional regularization, \be \phi\!\left(B_+^c
[X]\right)\left\{ \frac{q^2}{\mu^2};z\right\}\! =
[\mu^2]^{z\frac{\vert c\vert}{2}}\!\int\!\!  \, \frac{f_c(\vert y
\vert)\phi \left( X\right) \left\{ \frac{y^2}{\mu^2};z
\right\}}{y^2 + q^2} \, [y^2]^{-z(\frac{\vert c\vert }{2}-1) }d^Dy
\, , \label{toy}\ee for some functions $f_c(y)$ which turn to a
constant for $\vert y\vert\to\infty$. In the following, we assume
that $f_c$ is simply a constant, in which case the above Feynman
rules are elementary to compute. In the above, $ \phi
\left(T\right)$ is a function of a dimensionless variable
$q^2/\mu^2$ and the regularization parameter $z=(2-D)/2$.

   Furthermore \be
  \phi [X_1 \, X_2] =  \phi (X_1) \, \phi (X_2)
\ee as part of the definition and therefore $\phi [e] =  1$.
Hence, indeed, $\phi \in G$. We also provided for each decoration
$c$ an integer degree $\vert c\vert\geq 1$, which resembles the
loop number of skeleton diagrams.

The sole purpose of this choice of $\phi\in G$ for the Feynman
rules is to provide a simple character which suffers from
short-distance singularities in quite the same way as genuine
Feynman diagrams do, without confronting the reader with overly
hard analytic challenges at this moment.
 Note that, $\forall \, c \in {\rm Dec}$,
\be \phi \left( \treea{c}\; \right)  =  f_c \left[
\frac{q^2}{\mu^2} \right]^{\frac{-z\vert c \vert}{2}} \pi^{D/2} \,
\frac{\Gamma (1+\vert c\vert z)}{\vert c\vert z}  \, , \nonumber
\ee exhibiting the obvious pole at $D= 2$.

Using \be \int d^D y \, \frac{[y^2]^{-u}}{y^2 + q^2} = \pi^{D/2}
\, [q^2]^{-z-u}  \frac{\Gamma (-u + D/2) \, \Gamma (1+u -
D/2)}{\Gamma (D/2)} \, , \ee evaluations of decorated rooted trees
are indeed elementary. The reader can convince himself that the
degree of the highest order pole of $\phi(T)$ equals the
augmentation degree ${\rm aug}(T)$ of $T$, which, for a single
tree, is the number of vertices, see (\ref{aug}) below.

Having defined the character $\phi$, we note that, for $T = B_+^c
(X)$ \be \phi \circ S [B_+^c (X)] =  - \phi (T) - m \circ [\phi
\circ\! S\!\circ\!P \otimes \phi\!\circ\!B_+^c] \circ \Delta [X]
\, . \ee We then twist $\phi \circ S \in G$ to $S_R^{\phi} \in G$
by
\begin{eqnarray}
S_R^{\phi} (T) &:=  &-R [\phi (T) + m \circ (S_R^{\phi}\!\circ\!P
\otimes \phi \circ
B_+^c) \, \Delta [X]] \, ,  \label{SR}\\
&= : &-R [\bar\phi_R (T)]
\end{eqnarray}
where the renormalization scheme $R : V \rightarrow V$ is a
Rota--Baxter map and hence fulfills $R[ab] + R[a] \, R[b] = R
[R(a)b] + R[aR(b)]$, which suffices \cite{chen} to guarantee that
$S_R^{\phi} \in G$, as it guarantees that $S_R^\phi\circ
m_H=m_V\circ (S_R^\phi\otimes S_R^\phi)$.

   Set \be G \ni\phi_R (T) \equiv  S_R^{\phi} * \phi (T)
\equiv m \circ (S_R^{\phi} \otimes \phi) \circ \Delta [T] \, . \ee
Furthermore, assume that $R$ is chosen such that \be \lim_{z
\rightarrow 0} \ \left(\phi (X) - R [\phi (X)]\right) \quad
\hbox{exists} \ \forall \, X \in H \, .\label{assum} \ee
\subsection{Renormalizability and Hochschild Cohomology}
We now can prove renormalization for the Hopf algebra $H$ and the
toy Feynman rules $\phi$ in a manner which allows for a
straightforward generalization to QFT. \begin{theorem} i)
$\build\lim_{z \rightarrow 0}^{} \phi_R (T) \left\{
\frac{q^2}{\mu^2} ; z \right\}$ exists and is a polynomial in
$\log \frac{q^2}{\mu^2}$ (``finiteness'')\\ ii) $\build\lim_{z
\rightarrow 0}^{} \frac{\partial}{\partial \log q^2/\mu^2} \,
\bar\phi_R (T)\left\{ \frac{q^2}{\mu^2};z\right\}$ exists (``local
counterterms'').\end{theorem} To prove this theorem, we use that
$B_+^c$ is a Hochschild closed one-cocycle $\forall \, c \in {\rm
Dec}$.\\
{\it Proof.} For us,  Hochschild closedness just states \cite{CK1}
that \be \Delta \circ B_+^c (X) =  B_+^c (X) \otimes e + ({\rm id}
\otimes B_+^c) \, \Delta (X) \;\Leftrightarrow \;b \, B_+^c =  0
\, . \ee We want to prove the theorem in a way which goes through
unmodified in the context of genuine field theories. That
essentially demands that we only use Hopf algebra properties which
are true regardless of the chosen character representing the
Feynman rules. To this end
 we introduce the augmentation degree. Let $P$ be the projection
into the augmentation ideal, as before.

   Define, $\forall k\geq 2$, \be {\mathcal P}^k : H \rightarrow \underbrace{H
\otimes \ldots \otimes H}_{k \, {\rm copies}} \ee by \be [P
\otimes \ldots \otimes P] \circ \Delta^{k-1} \, , \ {\mathcal P}^1
:=  P
 \, ,
\ {\mathcal P}^0 :=  {\rm id} \, . \ee For every element $X$ in
$H$, there exists a largest  integer $k$ such that ${\mathcal P}^k
(X) \ne 0$, ${\mathcal P}^{k+1} (X) =  0$. We set \be{\rm aug} \,
[X] = k.\label{aug}\ee (This degree is  called bidegree in
\cite{BKprim}.) We prove the theorem by induction over this
augmentation degree. It suffices to prove it for trees $T \in
H_L$.\\
Start of the induction: ${\rm aug}(T)= 1$.\\
Then, $T =  \treea{c}\;$ for some $c \in {\rm Dec}$. Indeed \be
{\mathcal P}^1 (\treea{c}\;) \ne 0 \, , \ {\mathcal P}^2 (T) = ( P
\otimes P) [ \treea{c}\;\otimes e + e \otimes \treea{c}\; ] =  0
\, . \ee \be S_R^{\phi} (\treea{c}\;) =  -R [\phi (\treea{c}\;)]
\, , \ee and \be S_R^{\phi} * \phi (\treea{c}\;) =
\phi(\treea{c}\;)-R[\phi(\treea{c}\;)]\, , \ee which is finite by
assumption (\ref{assum}). Furthermore, $\lim_{z\to
0}\partial/\partial\log(q^2/\mu^2)\phi(\treea{c}\;)$ exists
$\forall c$, so
we obtain a start of the induction.\\
Induction: Now, assume that $\forall \, T$ up to ${\rm aug} \, (T)
=  k$, we have that \be \lim_{z \rightarrow 0} \
\frac{\partial}{\partial \log \frac{q^2}{\mu^2}} \, \bar\phi (T)
\ee exists, and $S_R^{\phi} * \phi (T)$ is a finite polynomial in
$\log \frac{q^2}{\mu^2}$ at $z= 0$. We want to prove the
corresponding properties for $T$ with ${\rm aug} \, (T) =  k+1$.

   So, consider $T$ with ${\rm aug} \, (T) =  k+1$.
Necessarily (each $T$ is in the image of some $B_+^c$), $T = B_+^c
(X)$ for some $c\in {\rm Dec}$ and $X\in H$. Then, from \be \Delta
\circ B_+^c (X) =  B_+^c (X) \otimes e + ({\rm id} \otimes B_+^c)
\, \Delta [X] \, , \ee indeed the very fact that $B_+^c$ is
Hochschild closed,  we get {\begin{eqnarray}  & & S_R^{\phi}
\left(B_+^c [X]\right)\left\{ \frac{q^2}{\mu^2};z\right\}  =  -R
\biggl[ \int \frac{(y^2)^{-(\frac{\vert c \vert}{2}-1) z}d^D
y}{[\mu^2]^{-\frac{\vert c\vert}{2} z}} \, \frac{f_c}{y^2 + q^2}
\, \phi (X) \left\{ \frac{y^2}{\mu^2} ; z \right\}
  \nonumber\\
  & & + \sum \int \frac{(y^2)^{-(\frac{\vert c \vert}{2}-1) z}d^D
y}{[\mu^2]^{-\frac{\vert c\vert}{2} z}} \, \frac{f_c}{y^2 + q^2}
\, S_R^{\phi} (X') \, \phi (X'') \left\{ \frac{y^2}{\mu^2} ; z
\right\} \biggl] \, ,
\end{eqnarray}}
where we abbreviated $\Delta [X] =  \sum X' \otimes X''$, and the
above can be written, using the definition (\ref{SR}) of
$S_R^\phi$, as { \bea  & & S_R^{\phi} \left(B_+^c
[X]\right)\left\{ \frac{q^2}{\mu^2};z\right\}
 =  \nonumber\\  & &  -R \left[ \int \frac{(y^2)^{-(\frac{\vert
c \vert}{2}-1) z}d^D y}{[\mu^2]^{-\frac{\vert c \vert}{2} z}} \,
\frac{f_c}{y^2 + q^2} \, S_R^{\phi} * \phi (X) \left\{
\frac{y^2}{\mu^2} ; z \right\} \right] \, .   \eea} This is the
crucial step: the counterterm is obtained by replacing the
subdivergences in $\phi(B_+^c(X))$ by their renormalized
evaluation $S_R^\phi\star\phi(X)$, thanks to the fact that
$bB_+^c=0$.

Now use that  ${\rm aug} \, [X] =  k$, and that $X$ is a product
$X \equiv \ \build\prod_{i}^{} \tilde T_i$ say, so that \be
S_R^{\phi}
* \phi (X) = \prod_i S_R^{\phi} * \phi (\tilde T_i) \, . \ee We
can apply the assumption of the induction to $S_R^{\phi} * \phi
(X)$. Hence there exists an integer $r_X$ such that  \be
S_R^\phi*\phi(X)\left\{ \frac{y^2}{\mu^2};z\right\}=
\sum_{j=0}^{r_X} c_j(z) [\log(y^2/\mu^2)]^j\ee for some
coefficient functions $c_j(z)$ which are regular at $z=0$.

A simple derivative with respect to $\log \frac{q^2}{\mu^2}$ shows
that $\bar\phi_R (T)$ has a limit when $z \rightarrow 0$ which
proves locality of $S_R^\phi$. Here, we use that our integrands
belong to the class of functions analyzed in \cite{W}. The needed
results for $S_R^{\phi}
* \phi (T)$ follow similarly. \hfill $\Box$\\ We encourage the
reader to go through these steps for a rooted tree with
augmentation degree three say.
\section{DSEs and factorization}
We start by considering  combinatorial DSEs. Those we define to be
equations which define formal series over Hopf algebra elements.
As before, we consider a Hopf algebra of decorated rooted trees,
with the corresponding investigation of DSEs in the Hopf algebra
of graphs to be given in \cite{DKnew}.
\subsection{The general structure of DSEs}
In analogy to the situation in QFT, our toy DSE considered here is
of the form \be X =  1 +  \sum_{c\in S\subseteq {\rm Dec}}
\alpha^{\vert c \vert -1 } B_+^{c} [X^{\vert c \vert}],
\label{X}\ee where $\forall \;c \in {\rm Dec}$, $\vert c \vert$ is
an integer chosen $\geq 2$, and the above is a series in $\alpha$
with coefficients in $H\equiv H(S)$. Note that every non-trivial
term on the rhs is in the image of a closed Hochschild one-cocycle
$B_+^{c}$.  The above becomes a series, \be X=1+\sum_{k=2}^\infty
c_k \alpha^{k-1}, \; c_k\in H, \ee  such that $c_k$ is a weighted
sum of all decorated trees with weight $k$. Here, the weight
$\vert T\vert$ of a rooted tree $T$ is defined as the sum of the
weights of its decorations: \be \vert T\vert:=\sum_{v\in
T^{[0]}}\vert{\rm dec}(v)\vert.\ee

This is typical for a Dyson--Schwinger equation, emphasizing the
dual role of the Hochschild one-cocycles $B_+^c$: their Hochschild
closedness guarantees locality of counterterms, and they define
quantum equations of motion at the same time. In the above,
$\alpha$ plays the role corresponding to a  coupling constant and
provides a suitable grading of trees by their weight.

Let us now assign to a given unordered set $I\subset {\rm Dec}$ of
decorations the linear combination of rooted trees \be
\underline{T}(I):=\sum_{T\in H \atop I= {\bigcup_{v\in
T^{[0]}}}{\rm dec}(v) }\frac{\alpha^{\vert T\vert -1}c_T}{{\rm
sym}(T)}T.\ee Here, the symmetry factor of a tree $T$ \cite{chen}
is the rank of its automorphism group, for example \be {\rm
sym}(\;\treecb{b}{a}{a}\;)=2,\;{\rm
sym}(\;\treecb{b}{b}{a}\;)=1.\ee To define $c_T$, let for each
vertex $v$ in a rooted tree $f_v$ be the number of outgoing edges
as in \cite{chen}. Then \be c_T:=\prod_{v\in T^{[0]}}\frac{\vert
{\rm dec}(v)\vert! }{(\vert{\rm dec}(v)\vert-f_v)!}.\ee If a tree
$T$ appears in such a sum, we write $T\in \underline{T}(I)$. It is
then easy to see that for such a linear combination $
\underline{T}(I)$ of rooted trees we can recover $I$ from ${\cal
P}^{{\rm aug}(T)}(T)$.  For two sets $I_1,_2$ we then define \be
\underline{T}(I_1)\vee\underline{T}(I_2):=\underline{T}(I_1\cup
I_2). \ee
\begin{theorem} For the DSE above, we have\\
 i) $X=1+\sum_{T\in H(S)}\alpha^{\vert T\vert -1}\frac{c_T}{{\rm sym}(T)} T,$\\
ii) $\Delta(c_k)=\sum_{i=0}^k{\rm Pol}_i\otimes c_{k-i},$ where $
{\rm Pol}_i$ is a degree $i$ polynomial in the $c_j$.
Thus, these coefficients $c_j$  form a closed subcoalgebra.\\
iii) $X=\prod_{c\in S}^\vee\frac{1}{1-\alpha^{\vert c \vert -1
}\underline{T}(c)}.$ The solution factorizes in terms of geometric
series with respect to the product $\vee$.
\end{theorem}
This theorem is a special case of a result in \cite{DKnew}, to
which we have to refer the reader for a proof. The factorization
in the third assertion is a triviality thanks to the definition of
$\underline{T}$. It only becomes interesting in the QFT case where
the pre-Lie product of graphs is degenerate \cite{banz}.
\subsection{Example}
To have a concrete example at hand, we focus on the equation: \be
X=1+\alpha B_+^a(X^2)+\alpha^2 B_+^b(X^3), \ee where we have
chosen  $\vert a\vert=2$ and $\vert b\vert=3$. For the first few
terms the expansions of $X$ reads
\begin{eqnarray}
c_1 & = & \treea{a}\;\;,\\
c_2 & = & \treea{b}+2\treeb{a}{a}\;\;,\\
 & & \nonumber\\
c_3 & = & 2\;\treeb{a}{b}\;
+3\;\treeb{b}{a}\;+4\;\treeca{a}{a}{a}\;+\;\treecb{a}{a}{a}\;\;,\\
 & & \nonumber\\
c_4 & = &
3\;\treeb{b}{b}\;+4\;\treeca{a}{a}{b}\;+6\;\treeca{a}{b}{a}\;+6\;\treeca{b}{a}{a}\;
+2\;\treecb{a}{a}{b}\;+3\;\treecb{b}{a}{a}\;+8\;\treeda{a}{a}{a}{a}\nonumber\\
 & & \nonumber\\
  & & +4\;\;\;\treedb{a}{a}{a}{a}\;\;
+2\;\treedc{a}{a}{a}{a}\;\;.
\end{eqnarray}
As rooted trees, we have non-planar decorated rooted trees, with
vertex fertility bounded by three in this example. In general, in
the Hopf algebra of decorated rooted trees, the trees with vertex
fertility $\leq k$, always  form a sub Hopf algebra.

Let us calculate the coproducts of $c_i$, $i=1,\dots 4$ say, to
check the second assertion of the theorem. We confirm \bea
\Delta(c_1) & = & c_1\otimes e+e\otimes c_1,\\
\Delta(c_2) & = & c_2\otimes e+e\otimes c_2+2c_1\otimes c_1,\\
\Delta(c_3) & = & c_3\otimes e+e\otimes c_3+3c_1\otimes c_2+[2c_2+c_1c_1]\otimes c_1,\\
\Delta(c_4) & = & c_4\otimes e+e\otimes c_4+4c_1\otimes
c_3+[3c_2+3c_1c_1]\otimes c_2\nonumber\\ & &
+[2c_3+2c_1c_2]\otimes c_1. \eea
\subsection{Analytic Factorization}
The crucial question now is what has the evaluation of all the
terms in $X$ as given by (\ref{X}), \be
\phi(X)\left\{\frac{q^2}{\mu^2};\alpha;z\right\}=1+\sum_{T\in
H(S)} \frac{c_T\alpha^{\vert T\vert -1 }}{{\rm sym}(T)}
\phi(T)\left\{\frac{q^2}{\mu^2};z\right\}\;,\ee to do with \be
\prod_{c\in S}\frac{1}{ 1-\alpha^{\vert c\vert -1
}\phi(\;\treea{c}\;)\left\{\frac{q^2}{\mu^2};z\right\} }?\ee If
the evaluation of a tree would decompose into the evaluation of
its decorations, we could expect a factorization of the form \be
\phi\left(\underline{T}(I)\right)=N_I\prod_{c\in
I}\phi(\treea{c}\;),\ee where $N_I$ is the integer $\sum_{T\in
\underline{T}(I)}c_T$. It is easy to see that the highest order
pole terms at each order of $\alpha$ in the unrenormalized DSE are
in accordance with such a factorization \cite{DK}, but that we do
not get a factorization for the non-leading terms.

>From the definition (\ref{toy})  for our toy model Feynman rule
$\phi$ we can write the DSE for the unrenormalized toy Green
function $\phi(X)$ as \be
\phi(X)\left\{\frac{q^2}{\mu^2};\alpha;z\right\}\!=1\!+\!\!\sum_{c\in
S}\alpha^{\vert c\vert -1}\int\! d^D\!y \frac{[y^2]^{z(\frac{\vert
c\vert }{2}-1) }f_c}{y^2+q^2}\phi(X)^{\vert c
\vert}\left\{\frac{y^2}{\mu^2};\alpha;z\right\}.\ee As the $B_+^c$
in (\ref{X}) are Hochschild closed, the corresponding renormalized
DSE is indeed of the same form \be
\phi_R(X)\!\!\left\{\!\frac{q^2}{\mu^2};\alpha;z\!\right\}\!\!=\!\!Z_X\!+\!\!\sum_{c\in
S}\!\!\alpha^{\vert c\vert -1}\!\int\!\! d^D\!y
\frac{[y^2]^{z(\frac{\vert c\vert }{2}-1)
}f_c}{y^2+q^2}\phi_R(X)^{\vert c
\vert}\left\{\frac{y^2}{\mu^2};\alpha;\!z\!\!\right\},\ee where
$Z_X=S_R^\phi(X)$.

Now assume we would have some "RG-type" information about the
asymptotic behaviour of $\phi_R(X)$, for example \be
\phi_R(X)\left\{\frac{q^2}{\mu^2};\alpha\right\}=F(X)(\alpha)\left[\frac{q^2}{\mu^2}\right]^{-\gamma(\alpha)}
,\ee consistent with the renormalized DSE. Then, our toy model
would regulate itself, as \bea
\phi_R(X)\left\{\frac{q^2}{\mu^2};\alpha\right\} & = &
[\mu^2]^{\gamma(\alpha)}\sum_{c\in S
}\alpha^{\vert c \vert -1}\nonumber\\
 & \times & \int
d^2y \frac{[y^2]^{(\vert c \vert -1
)\gamma(\alpha)}f_c}{y^2+q^2} \left[
\phi_R(X)\left\{\frac{y^2}{\mu^2},\alpha\right\}\right]^{\vert c \vert}\\
 & = &
[\mu^2]^{\gamma(\alpha)}\sum_{c\in S }\alpha^{\vert c \vert -1}
\left[F(X)(\alpha)\right]^{\vert c\vert}\;\int d^2y
\frac{[y^2]^{-\gamma(\alpha)}f_c}{y^2+q^2}\; ,\label{fact}\eea
with no need for a regulator, as long as we assume that
$\gamma(\alpha)$ serves that purpose, possibly by means of
analytic continuation.

Then, we would be in much better shape: the "toy anomalous
dimension" $\gamma(\alpha)$ could be defined from the study of
scaling in the complex Lie algebra ${\cal L}$ underlying the dual
of $H(S)$ \cite{RH2} while $F(X)(\alpha)$ could be recursively
determined at $q^2=\mu^2$ from Feynman rules which imply
factorization for a tree $T=B_+(U)$ as \be
\phi_R(B_+^c(U))\left\{\frac{q^2}{\mu^2};\alpha\right\}=\phi_R(\treea{c}\;)\left\{\frac{q^2}{\mu^2};\alpha\right\}\;\phi(U)\left\{1;\alpha\right\},\ee
by (\ref{fact}).

Alas, we do not have a renormalization group at our disposal here.
But in QFT we do. While it might not tell us that we have scaling
\cite{SC}, it will indeed give us information about the asymptotic
behaviour, which combines with the present analysis of DSEs in a
profitable manner: what is needed is information how the
asymptotic behaviour of the integrand which corresponds to $B_+^c$
under the Feynman rules relates to the asymptotic behaviour of the
integral. This is just what field theory provides. We will in
\cite{DKnew} then indeed  set out to combine the DSEs and the RG
so as to achieve a factorization in terms of Hopf algebra
primitives, using the Hochschild closedness of suitable
$B_+^\gamma$ operators, the RG, as well as a dedicated choice of
Hopf algebra primitives so as to isolate all short-distance
singularities in Green functions which depend only on a single
scale. As it will turn out, this makes the Riemann--Hilbert
approach of \cite{RH1,RH2} much more powerful.
\subsection{Remarks}
Let us understand how the above theorem fares in the context of
QFT. Consider all 1PI graphs together with their canonical Hopf-
and Lie algebra structures of 1PI graphs. The set of primitive
graphs is then well-defined. We use it to form a set of equations
\be\Gamma^{\underline{n}}=1+ \sum_{\gamma\in H^{[1]}_L \atop {\rm
res}(\gamma)=\underline{n} }\frac{g^{\vert\gamma\vert-1}}{{\rm
Sym}(\gamma)}B_+^\gamma (X_{\cal R}^\gamma).\ee These equations
define  1PI Green functions, in a normalization such that its tree
level value is unity, recursively, via insertion of such Green
functions (combined in a monomial $X_{\cal R}^\gamma$) into prime
graphs $\gamma$, graphs which are themselves free of subgraphs
which are superficially divergent. They define  formal series in
graphs such that the evaluation by the Feynman rules delivers the
usual quantum equations of motion, the DSEs. This gives us an
independent way to find such equations of motion: the above
equation can be described as a canonical problem in Hochschild
cohomology, without any reference to the underlying physics.
Investigating these equations from that viewpoint has many
interesting consequences \cite{DKnew} which generalize the toy
analysis in this talk: \begin{enumerate} \item The
$\Gamma^{\underline{n}}$ are determined as the sum over all 1PI
graphs with the right weights so as to determine the 1PI Green
functions of the theory:
\be\Gamma^{\underline{n}}=1+\sum_{\Gamma\in H_L\atop {\rm
res}(\gamma)=\underline{n}}\frac{g^{\vert\Gamma\vert}}{{\rm
Sym}(\Gamma)}\Gamma, \ee where the latter sum is over all 1PI
graphs $\Gamma$ with external legs ("residue") $ \underline{n}$.
\item The maps $B_+^\gamma$ are suitably defined
so that they are Hochschild closed for a sub Hopf algebra of
saturated sums of graphs $\Sigma_\Gamma=\sum_i \gamma_i\star X_i$
which contain all maximal forests: \be \sum_i \Delta
B_+^{\gamma_i}(X_i)=\sum_i B_+^{\gamma_i}(X_i)\otimes e+\sum_i
({\rm id}\otimes  B_+^{\gamma_i})\Delta(X_i)\; . \ee
\item This delivers a general proof of locality of counterterms
and finiteness of renormalized Green functions by induction over
the augmentation degree precisely as above:  \bea \sum_i \Delta
B_+^{\gamma_i}(X_i) & = & \sum_i  B_+^{\gamma_i}(X_i)\otimes
e+\sum_i ({\rm
id}\otimes B_+^{\gamma_i})\Delta(X_i)\;\Leftrightarrow\\
 & & \sum_i S_R^\phi(B_+^{\gamma_i}(X_i))=({\rm id}-R)\sum_i \int
D(\gamma)(S_R^\phi\star\phi(X_i)),\nonumber\eea so that the
$R$-bar operation and the counterterm are obtained by replacing
the divergent subgraphs by their renormalized contribution.

\item The terms of a given order in a 1PI Green functions form a
closed Hopf subalgebra: \be \Gamma^{\underline{m}}=:1+\sum_k
c_k^{\underline{m}}\;g^k\Rightarrow
\;\Delta(c_k^{\underline{m}})=\sum_{j=0}^k {\rm
Pol}^{\underline{m}}_{j}\otimes c_{k-j}^{\underline{m}}\;,\ee
where the $ {\rm Pol}^{\underline{m}}_{j}$ are monomials in the
$c_j^{\underline{n}}$ of degree $j$, where $\underline{n}\in {\cal
R}$.  Thus, the space of polynomials in the $c_k^{\underline{m}}$
is a closed Hopf sub(co)algebra of $H$. This is a subtle surprise:
to get this result, it is necessary and sufficient to impose
relations between Hopf algebra elements: \be \forall
\gamma_1,\gamma_2\in c_1^{\underline{n}}, \;X_{\cal
R}^{\gamma_1}=X_{\cal R}^{\gamma_2}.\ee These relations turn out
to be good old friends, reflecting the quantum gauge symmetries of
the theory: they describe the kernel of the characters
$\phi,S^\phi_R, S_R^\phi\star\phi$, and translate to the
Slavnov--Taylor identities \be
\frac{Z^{\qcdfv}}{Z^{\qedf}}=\frac{Z^{\qcdgc}}{Z^{\qcdgb}}=\frac{Z^{\qcdgd}}{Z^{\qcdgc}}
=\frac{Z^{\qcdghv}}{Z^{\qcdgh}}\; , \ee where
$Z^{\ldots}=S_R^\phi(\Gamma^{\ldots})$.
\item The effective action, as a sum over all 1PI Green functions,
factorizes uniquely into prime graphs with respect to a
commutative associative product on 1PI graphs $\vee$: \be
S^\Gamma_{eff}=\sum_{\underline{m}}\Gamma^{\underline{m}}
=\prod^\vee_{\gamma\in H_L^{[1]}
}\frac{1}{1-g^{\vert\gamma\vert-1}\underline{\Gamma}(\gamma)}.\ee
Integrality of this product again relates back to relations
between graphs which correspond to Ward identities.
\end{enumerate}
With these remarks, we close and invite the reader to participate
in the still exciting endeavour to understand the structure of
renormalizable quantum field theories in four dimensions.
\subsubsection*{Acknowledgments} {\small It is a pleasure to thank
participants and organizers of our school for a wonderful (and
everywhere dense) atmosphere. thanks to K.~Ebrahimi-Fard for
proofreading the ms. This work was supported in parts by NSF grant
DMS-0205977 at the Center for Mathematical Physics at Boston
University.}

%

\printindex
\end{document}